\documentclass[superscriptaddress,twocolumn,showpacs,preprintnumbers,amsmath,amssymb,prb]{revtex4}

\usepackage{graphicx}
\usepackage{dcolumn}
\usepackage{bm}

\begin{document}

\preprint{APS/123-QED}

\title{Orbital selectivity of the kink in the dispersion of Sr$_{2}$RuO$_{4}$}

\author{H. Iwasawa}
\affiliation{Department of Applied Physics, Tokyo University of Science, Shinjuku-ku, Tokyo 162-8601, Japan}
\affiliation{National Institute of Advanced Industrial Science and Technology, Tsukuba, Ibaraki 305-8568, Japan}

\author{Y. Aiura}
\email[To whom all correspondence should be addressed.\\ Electronic address: ]{y.aiura@aist.go.jp}
\affiliation{National Institute of Advanced Industrial Science and Technology, Tsukuba, Ibaraki 305-8568, Japan}

\author{T. Saitoh}
\affiliation{Department of Applied Physics, Tokyo University of Science, Shinjuku-ku, Tokyo 162-8601, Japan}

\author{I. Hase}
\author{S. I. Ikeda}
\author{Y. Yoshida}
\author{H. Bando}
\affiliation{National Institute of Advanced Industrial Science and Technology, Tsukuba, Ibaraki 305-8568, Japan}

\author{M. Higashiguchi}
\author{Y. Miura}
\author{X.Y. Cui}
\affiliation{Graduate School of Science, Hiroshima University, Higashi-Hiroshima 739-8526, Japan}

\author{K. Shimada}
\author{H. Namatame}
\affiliation{Hiroshima Synchrotron Radiation Center, Hiroshima University, Higashi-Hiroshima 739-8526, Japan}

\author{M. Taniguchi}
\affiliation{Graduate School of Science, Hiroshima University, Higashi-Hiroshima 739-8526, Japan}
\affiliation{Hiroshima Synchrotron Radiation Center, Hiroshima University, Higashi-Hiroshima 739-8526, Japan}

\date{\today}

\begin{abstract}

We present detailed energy dispersions near the Fermi level on the monolayer perovskite ruthenate Sr$_{2}$RuO$_{4}$, determined by high-resolution angle-resolved photoemission spectroscopy.
An orbital selectivity of the kink in the dispersion of Sr$_{2}$RuO$_{4}$ has been found: A kink for the Ru $4d_{xy}$ orbital is clearly observed, but not for the Ru $4d_{yz}$ and $4d_{zx}$ ones.
The result provides insight into the origin of the kink.

\end{abstract}


\maketitle

% ********** INTRODUCTION **********
\section{INTRODUCTION}
The sudden change of the group velocity, so-called ``kink", of the dispersing peak in angle-resolved photoemission spectroscopy (ARPES) spectra is widely reported in the high-temperature superconducting cuprates.\cite{Kaminski01, Johnson01, Lanzara01, Shen02, Zhou03}
Nevertheless, the interpretations of the kink have been controversial.
In previous works, electronic coupling to a bosonic mode such as phonons\cite{Lanzara01, Shen02, Gweon04} or magnetic excitations\cite{Scalapino99, Carbotte99, He02, Norman04} has been discussed as its origin.
Recently, we have found a similar kink of the layered strontium ruthenates,\cite{Aiura04} meaning that the kink is not peculiar to the cuprates.
The layered ruthenates are isostructural to the cuprates, while the electronic and magnetic properties are quite different.
The electronic structure close to the Fermi level ($E_{\rm F}$) of the layered ruthenates is derived not only from the in-plane Ru $4d_{xy}$$-$O $2p$ band but also from the out-of-plane Ru 4$d_{yz,zx}$$-$O 2$p$ ones, while for the cuprates the single in-plane Cu 3$d_{x^{2}-y^{2}}$$-$O 2$p$ band plays a crucial role.
Therefore, ARPES study on the layered ruthenates is expected to provide insight into the origin of the kink in transition-metal oxides.

The monolayer ruthenate Sr$_{2}$RuO$_{4}$ have attracted significant attention since the spin-triplet superconductivity was discovered below $T_{\rm c}$=1.5 K.\cite{Maeno94, Ishida98, Maeno01}
The electronic bands near $E_{\rm F}$ are derived mainly from the Ru $t_{2g}$ orbitals; the band near $E_{\rm F}$ due to the in-plane Ru $4d_{xy}$ orbital exhibits two-dimensional (2D) character with a large energy dispersion, while the bands due to the out-of-plane Ru 4$d_{yz}$ and 4$d_{zx}$ orbitals exhibit nearly 1D character with a small energy dispersion.
Previous ARPES and de Haas-van Alphen (dHvA) studies showed the Fermi surface (FS) exhibiting one hole sheet ($\alpha$) and two electron sheets ($\beta$ and $\gamma$),\cite{Mackenzie96, Damascelli00} qualitatively consistent with the band-structure calculation based on the local density approximation (LDA).\cite{Oguchi95, Singh95, Hase96, Mazin97}
It should be noted that except for the vicinity of the high symmetry line ${\Gamma}X$, the $\alpha$ and $\beta$ sheets are derived from the Ru $4d_{yz}$ and $4d_{zx}$ orbitals, while the remaining $\gamma$ sheet is from the Ru $4d_{xy}$ orbital.
Moreover, along the ${\Gamma}X$ line, it is difficult to assign the orbital character to each sheet because those bands approach one another towards $E_{\rm F}$ and the hybridization among the different orbitals becomes significant (see Fig.~\ref{FS}).
The experimental geometry in our previous work was this,\cite{Aiura04} and hence we could not observe any conclusive relationship between the kink and the orbital, although such information should provide a key to elucidate the origin of the kink.
In this paper, we present detailed band dispersion of Sr$_{2}$RuO$_{4}$ not along the ${\Gamma}X$ line but the ${\Gamma}M$ and $MX$ lines, determined by high-resolution ARPES, to elucidate the orbital selectivity of the kink.
The spectral intensity of the band near the $M$ point along the ${\Gamma}M$ line is drastically reduced with increasing the angle of the incidence of the light, indicating that the band has the in-plane $4d_{xy}$ orbital character. 
A kink in the dispersion is clearly shown for this {\em xy} band, while not for the ${yz}$ and ${zx}$ ones.

% ********** EXPERIMENTAL **********
\section{EXPERIMENTAL}
\begin{figure}
\includegraphics{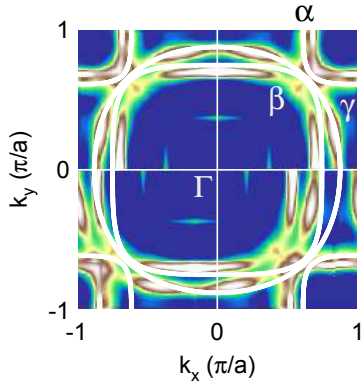}
\caption{(Color online). $E_{\rm F}$ intensity map taken with 65 eV at 30 K.
White lines denote the FSs based on the LDA calculation.\cite{Hase96}} 
\label{FS}
\end{figure}

High-quality Sr$_2$RuO$_4$ single crystals were grown by the floating zone method with a self-flux technique, resulting in a sharp superconducting transition at $T_{\rm c}$$\sim$1.36 K.\cite{Ikeda02,Yoshida99,Mao00} 
Figure.~\ref{FS} shows the FSs of Sr$_2$RuO$_4$, recorded at a high-resolution and high-flux undulator beamline (BL-28) of the Photon Factory.
Since the FSs are consistent with the LDA band-structure calculation\cite{Hase96} (white lines) and the previous ARPES result,\cite{Damascelli00} the quality of our samples should be high enough to investigate detailed band dispersions.
Except for the data in Fig.~\ref{FS}, all the ARPES measurements were carried out at BL-1 of Hiroshima Synchrotron Radiation Center (HSRC) in Hiroshima University,\cite{Shimada01, Shimada02} using 29 eV photons at 10 K.
The radiation is linearly polarized in the horizontal plane of incidence.
The sample goniometer provides independent polar and tilt rotations of the sample (R-Dec Co. Ltd., {\em i} GONIO LT).\cite{Aiura03b}
The beamline is equipped with a high-resolution, hemispherical electron analyzer (SCIENTA ESCA200). 
In order to obtain clean surfaces, we cleaved the samples {\em in situ} in ultrahigh vacuum better than $1 \times 10^{-10}$ Torr at 10 K.
A surface state near the $M$ point due to the rotation of RuO$_6$ octahedra at the surface was observed for the fresh surface,\cite{Matzdorf00, Shen01} but was almost eliminated by aging the sample surface {\em in situ}.\cite{Wang04}
The replica of FSs in the bulk due to the surface rotation was not shown on the aged surface used here, and the coherent peak dispersion and line shape were consistent with the dHvA results and the LDA band predictions.\cite{Mackenzie96, Oguchi95, Singh95, Hase96}
The angular resolution was $0.5^{\circ}$ (vertical) $\times 0.3^{\circ}$ (horizontal) and the spatial resolution of the angular window ($0.3^{\circ}$) corresponds to the $k$ resolution of 1.1 $\%$ of the ${\Gamma}M$ line. 
The total instrumental energy resolution was set to 20 meV.
The emission angle of the photoelectron with respect to the surface normal was varied by rotating the polar and tilt axes of the sample.

% ********** RESULT **********
\section{RESULT}

\begin{figure}
\includegraphics{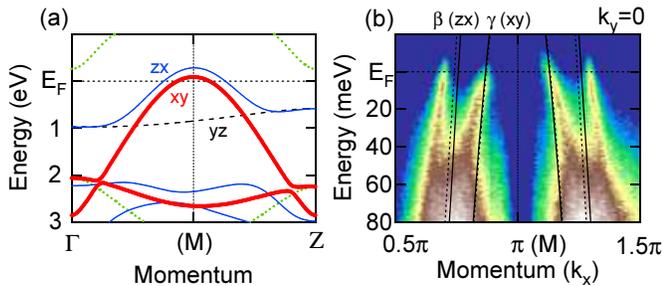}
\caption{(Color online). 
(a) The calculated energy bands of Sr$_{2}$RuO$_{4}$ along the ${\Gamma}Z$ line,\cite{Hase96} and (b) the intensity plot of the ARPES spectra along the (0.5$\pi$,0)-(1.5$\pi$,0) line. 
Each line style in (a) denotes the different irreducible expressions: $\Sigma_1$ (dotted), $\Sigma_2$ (dashed), $\Sigma_3$ (solid), $\Sigma_4$ (thick).}
\label{FIG2}
\end{figure}

Figure~\ref{FIG2} (a) shows the calculated energy bands of Sr$_2$RuO$_4$ along the ${\Gamma}MZ$ line near $E_{\rm F}$. 
There are two bands crossing $E_{\rm F}$: the band near the $M$ point is due to the $4d_{xy}$ orbital (thick line) and the other band far from the $M$ point is due to the $4d_{zx}$ one (solid line).\cite{comment1}
As seen in Fig.~\ref{FS}, the {\em xy} and {\em zx} bands compose the electron-like FS sheets $\gamma$ and $\beta$, respectively.
Along the $MX$ line, there is a band crossing $E_{\rm F}$ due to the $4d_{yz}$ orbital, which composes the hole-like $\alpha$ sheet.
Figure~\ref{FIG2} (b) shows the intensity plot of ARPES spectra along the (0.5$\pi$,0)-(1.5$\pi$,0) line together with the calculated dispersions of {\em zx} and {\em xy} bands.
At first sight, two prominent spectral features are clearly observed in both the first [(0,0)-($\pi$,0)] and the second [($\pi$,0)-(2$\pi$,0)] Brillouin zones.
Solid and dashed lines are the calculated dispersions along the ${\Gamma}MZ$ and $ZM{\Gamma}$ lines, respectively.
The slight difference between them represents the very small energy dispersion along the $k_z$ direction, indicating the strong 2D character of the electronic structure in Sr$_2$RuO$_4$.
Comparing the experiment with the calculation, one can suppose that the spectral feature adjacent to the $M$ point is derived from the {\em xy} band, and the other is from the {\em zx} band, although there exists an appreciable difference in the Fermi momentum ($k_{\rm F}$) between the calculation and experiment in the {\em zx} band. 

\begin{figure}
\includegraphics{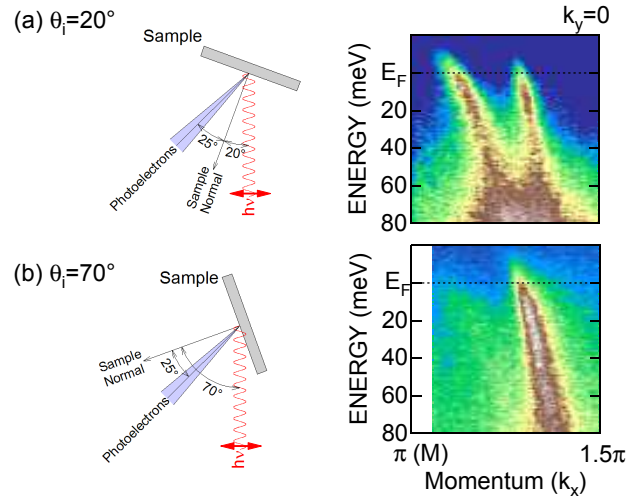}
\caption{(Color online). The experimental configuration and the intensity plot of Sr$_2$RuO$_4$ along the ($\pi$,0)-(1.5$\pi$,0) line for two $\theta_i$'s; (a) $\theta_i=20^\circ$ and (b) $\theta_i=70^\circ$.}
\label{FIG3}
\end{figure}

To determine precisely the orbital character of the two observed spectral features, we have selected two different angles of incidence ($\theta_i$) of the light.
Figure~\ref{FIG3} shows the experimental configurations and the corresponding intensity plots in the second Brillouin zone for different two $\theta_i$'s; the nearly (a) normal ($\theta_i$=$20^\circ$) and (b) parallel ($\theta_i$=$70^\circ$) with respect to the sample surface ({\em xy}-plane).
The spectral feature in the vicinity of the $M$ point becomes obscure as $\theta_i$ is increased.
According to the optical selection rule, the cross section of the Ru $4d_{yz}$ and $4d_{zx}$ orbitals is enhanced for the parallel photons, compared with that of the Ru 4$d_{xy}$ orbital.
Therefore, the drastic reduction of the spectral weight for the band near the $M$ point at the larger $\theta_i$ indicates that this band is derived from the Ru 4$d_{xy}$ orbital, while the other is derived from the Ru 4$d_{zx}$ orbital.
This assignment is consistent with the LDA calculations, as shown in Fig.~\ref{FIG2} (a).

\begin{figure}
\includegraphics{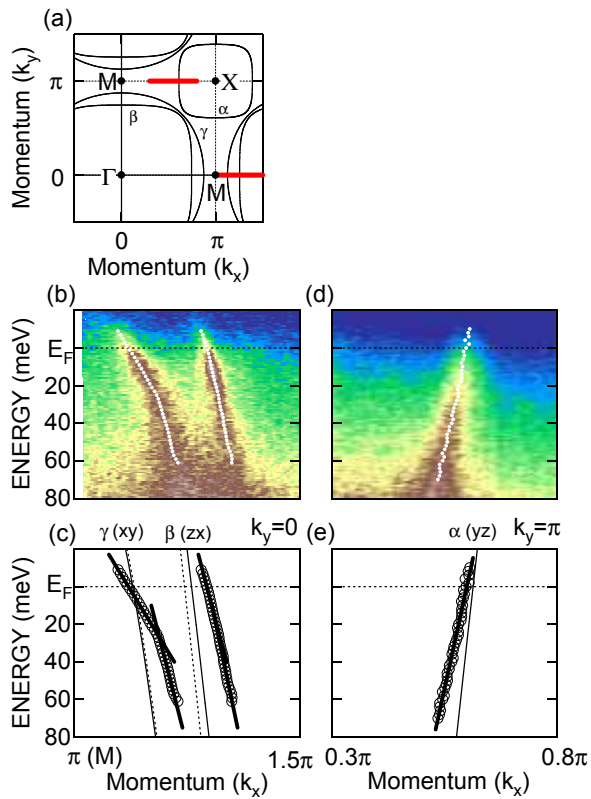}
\caption{(Color online). (a) FSs of Sr$_{2}$RuO$_{4}$ derived from the LDA calculations.  The measured momentum area in (b)-(e) are shown by the thick lines.
(b) and (c) Intensity plot of the ARPES spectra along the ($\pi$,0)-(1.5$\pi$,0) line and the MDC-derived dispersions (open circles) together with the calculated band structure. 
(d) and (e) Same as in (b) and (d) except the observed momentum area is now changed along the (0.3$\pi$,$\pi$)-(0.8$\pi$,$\pi$) line.}
\label{FIG4}
\end{figure}

Figure~\ref{FIG4} shows the intensity plot of the ARPES spectra along the ($\pi$,0)-(1.5$\pi$,0) and (0.3$\pi$,$\pi$)-(0.8$\pi$,$\pi$) lines and the dispersions (open circles) derived from the momentum distribution curve (MDC) together with the calculated dispersions of ${xy}$ and ${zx}$ bands.
The FSs and the measured momentum area (thick lines) are shown in Fig.~\ref{FIG4} (a).
In Fig.~\ref{FIG4} (b), one can observe a clear kink structure in the {\em xy} band, while such a kink is not observed in the {\em zx} band.
This observation is further confirmed by Fig.~\ref{FIG4} (c), where the dispersions (open circles) derived from the MDC is compared with the calculated dispersion of the {\em xy} and {\em zx} bands.
The band which crosses $E_{\rm F}$ in the (0.3$\pi$,$\pi$)-(0.8$\pi$,$\pi$) line and composes the $\alpha$ sheet, shows no kink [Figs.~\ref{FIG4} (d) and (e)].
In our previous study along the ${\Gamma}X$ line,\cite{Aiura04} the kink was observed in the $\alpha$ and/or $\gamma$ FS sheets derived from the out-of-plane Ru $4d_{yz}$ and $4d_{zx}$ orbitals, which seems to be contrary to the present case.
Except for near the ${\Gamma}X$ line, however, we confirmed by the FS mapping that the kink is clearly observed for $\gamma$ FS sheet derived from the in-plane $4d_{xy}$ orbital, while not for $\alpha$ and $\beta$ FS sheets derived from the out-of-plane Ru $4d_{yz}$ and $4d_{zx}$ orbitals.\cite{comment2}
Hence, we believe that the kink essentially exists only in $\gamma$ FS sheet out of the three FS sheets.
Here, we estimate the mass enhancement with respect to the band theory of the $\alpha$, $\beta$ and $\gamma$ FS sheets to be 2.0, 2.0 and 3.7, respectively. 
These values are qualitatively consistent with the cyclotron mass in the previous dHvA studies.\cite{Mackenzie96, Mackenzie03}
Consequently, the effective mass of the $\gamma$ FS sheet, which shows the kink, is substantially enhanced in contrast to those of the $\alpha$ and $\beta$ FS sheets.

% ********** DISCUSSION **********
\section{DISCUSSION}
%Coulomb correlations
As an origin of the observed mass enhancement, one can first raise the on-site Ru {\em d-d} Coulomb interaction $U$.
Previous photoemission studies have estimated $U$ to be 1.3$\sim$1.5 eV.\cite{Yokoya96, Inoue98}
On the other hand, the band theory predicts a roughly 3.5 eV wide in-plane {\em xy} band, and a narrow (1.4 eV width) out-of-plane {\em yz} and {\em zx} bands near $E_{\rm F}$.\cite{Hase96} 
Therefore, a larger mass enhancement should be expected for the {\em zx} band due to the substantial influence of the Coulomb correlations, rather than for the {\em xy} band, contrary to the ARPES results.
According to the perturbation theory and quantum Monte Carlo calculations for a multiband Hamiltonian,\cite{Liebsch 00} the narrow {\em yz} and {\em zx} bands are more strongly distorted by the Coulomb correlations than the wide {\em xy} band.
The charge is, then, partially transferred from the {\em yz} and {\em zx} bands to the {\em xy} band, leading that the Van Hove singularity of the {\em xy} band at the $M$ point shifts toward $E_{\rm F}$.
Based on this picture, it seems reasonable to suppose that the effective mass for the {\em xy} band is enhanced from the LDA band mass due to the charge transfer.
However, this should be accompanied by a shift of $k_{\rm F}$ for the {\em xy} band, which is not the case in the ARPES spectra.
Thus the mass enhancement for the {\em xy} band is probably caused not by the Coulomb correlations but by the kink as shown in Figs.~\ref{FIG4} (b) and (c).

%magnetic excitation
Then what is the origin of the kink of the {\em xy} band?
In recent works on the cuprates, the kink was related to bosonic modes such as phonons\cite{Lanzara01, Shen02, Gweon04} or magnetic excitations.\cite{Scalapino99, Carbotte99, He02, Norman04}
Manske, Eremin, and Bennemann\cite{Manske03} predicted a kink of Sr$_{2}$RuO$_{4}$ due to incommensurate antiferromagnetic spin fluctuations at ${\bf Q}_{ic}^{\alpha\beta}\approx(2\pi/3, 2\pi/3)$ based on the nesting properties of the quasi-1D $\alpha$ and $\beta$ bands.\cite{Mazin99, Sidis99, Morr01}
According to this prediction, the quasiparticles should be strongly renormalized due to coupling with the spin fluctuations for the $\alpha$ and $\beta$ bands.
However, the observed kink was not in the $\alpha$ nor $\beta$ bands but in the $\gamma$ band.
Recent neutron experiments revealed another weak and broadened magnetic structure around ${\bf Q}_{ic}^{\gamma}\approx(0.2\pi,0.2\pi)$, attributed to the excitation from the $\gamma$ band, in addition to the magnetic fluctuations at ${\bf Q}_{ic}^{\alpha\beta}$.\cite{Braden02, Kikugawa04}
Currently, we have no definite information on the relationship between the magnetic fluctuations at ${\bf Q}_{ic}^{\gamma}$ and the electronic structure presented here.
On the other hand, Matzdorf {\em et al.} suggested the strong coupling between the phonon mode corresponding to the in-plane octahedron rotation with the $\Sigma_{3}$ symmetry, which was shown by neutron experiments,\cite{Braden98} and ferromagnetic (FM) spin fluctuations.\cite{Matzdorf00} 
It is possible that the $\Sigma_{3}$ phonon mode couples strongly with the Ru 4$d_{xy}$ orbital but not with the Ru 4$d_{yz}$ and 4$d_{zx}$ orbitals, because the kink appears only in the {\em xy} band in our result.

%phonon excitation
Assuming the strong coupling between the electronic structure and the phonons, which is clearly indicated by the oxygen isotope effect on $T_{\rm c}$ of Sr$_{2}$RuO$_{4}$,\cite{Mao01} we discuss the origin of the kink shortly.\cite{comment3}
In a previous Raman study, the electron-phonon interaction for the apical oxygen vibration modes was observed.\cite{Sakita01}
However, it may be difficult to consider that the kink is caused by the effective coupling between the zone-center apical oxygen phonon modes and the in-plane {\em xy} band with the kink.
Directly relating to the in-plane Ru 4$d_{xy}$ orbital, there are four in-plane oxygen phonon modes at the zone-center for the tetragonal K$_{2}$NiF$_{4}$ structure; $B_{2u}$, $A_{2u}$, and two $E_{2u}$.
The $B_{2u}$ and $A_{2u}$ modes are the out-of-plane out-of-phase and out-of-plane in-phase vibrations of in-plane oxygen atoms, respectively.  
The remaining two $E_{2u}$ modes are related with the zone boundary in-plane oxygen-stretching longitudinal optical phonon, observed in the cuprates,\cite{Lanzara01} and the zone-boundary in-plane oxygen-rotation phonon\cite{Braden98} due to the FM fluctuations as mentioned above.\cite{Matzdorf00}
On the other hand, Devereaux {\em et al.} recently suggested for the bilayer cuprate Bi2212 that the out-of-plane, out-of-phase oxygen buckling mode ($B_{1g}$) couples strongly with the electronic states near the ($\pi$,0) point,\cite{Devereaux04} in agreement with ARPES spectra.\cite{Cuk04}
This $B_{1g}$ mode for the bilayer cuprate corresponds to the $B_{2u}$ mode for the monolayer Sr$_{2}$RuO$_{4}$.
In order to estimate the energy of the $B_{2u}$ mode, which is silent in Raman scattering, we calculated the phonon dispersion relation using the simple constant force model and the optical spectra.\cite{Udagawa98, Katsufuji96}
The obtained energy ($\sim$40 meV) is comparable with the binding energy of the kink of the {\em xy} band in Figs.~\ref{FIG4} (b) and (c).
Hence, we believe that the $E_{2u}$ and $B_{2u}$ modes are strong candidates to explain the kink, although more detailed studies on the phonon modes are needed.

% ********** CONCLUSION **********
\section{CONCLUSION}
In summary, we have found orbital selectivity of the kink in the dispersion of Sr$_{2}$RuO$_{4}$ by ARPES measurements along the ${\Gamma}M$ and $MX$ lines.
The kink is clearly observed for $\gamma$ FS sheet derived from the in-plane Ru $4d_{xy}$ orbital, while not for $\alpha$ and $\beta$ FS sheets derived from the out-of-plane Ru $4d_{yz}$ and $4d_{zx}$ orbitals, respectively.
The result gives critical restrictions for the origin of the kink.

\begin{acknowledgments}
One of the authors (Y. A.) thanks to H. Eisaki for useful discussion.
Y. A. and T. S. thank the users' working group for valuable help during the construction of the ARPES end station at the BL-28 of Photon Factory (KEK, Tsukuba).
The sample goniometer was partly developed under a Joint Development Research at High Energy Accelerator Research Organization (KEK, Tsukuba).
This work was partly supported by a Grant-in-Aid for COE Research (13CE2002) by the Ministry of Education, Science, and Culture of Japan.
We thank the Cryogenic Center, Hiroshima University for supplying liquid helium.
The synchrotron radiation experiments have been done under the approval of HSRC (Proposal No. 04-A-36).
\end{acknowledgments}

%\bibliography{prb}% Produces the bibliography via BibTeX.

\begin{thebibliography}{99}
\bibitem{comment1} The momentum size is defined so that the reciprocal vector at the midpoint between the $\Gamma$ and $Z$ points is scaled to $\pi$. That is, $\pi$ corresponds to $\pi$/$a$, where $a$ is the lattice constant. For convenience, the midpoint is called the $M$ point, which is used in simplified 2D Brillouin zone. Furthermore, it is assumed that the $x$ axis corresponds to the direction of the ${\Gamma}Z$ line. Symmetry symbols are taken from the notation used in the band calculation for the same body centered tetragonal structure La$_{2}$CuO$_{4}$; K. Takegahara, H. Harima, and A. Yanase, Jpn. J. Appl. Phys. {\bf26}, L352 (1987).
\bibitem{comment2} This situation is similar to the cuprate case in which the phonon modes relevant to the kink are different between along the node direction and the anti-node direction\cite{Lanzara01,Shen02, Devereaux04, Cuk04}.  Assuming that the kink is caused by the electron-phonon coupling, it is likely that the phonon vibration mode to couple the electronic structure along the ${\Gamma}X$  line is different from that along the ${\Gamma}M$ line.
\bibitem{comment3} We only consider the in-plane orbital character here because the kink is observed in the $\gamma$ band with in-plane character in the most of the Brillouin zone except for along the ${\Gamma}X$ line (see Ref.~\onlinecite{comment2}).
\end{thebibliography}

\end{document}